\documentclass[useAMS,usenatbib]{mn2e}
\usepackage{amsmath,fleqn,graphicx,amssymb}
\usepackage{multirow}
\renewcommand{\[}{\begin{equation}}
\renewcommand{\]}{\end{equation}}
\def\p{\partial}

\def\ex#1{\left\langle#1\right\rangle}

\let\boldgrk=\gkvecten
\let\boldgrksc=\gkvecseven

\def\gkthing#1{{\mathchoice%
	{\hbox{{\boldgrk\char#1}}}
	{\hbox{{\boldgrk\char#1}}}
	{\hbox{{\boldgrksc\char#1}}}
	{\hbox{{\boldgrksc\char#1}}}}}

\def\vdelta{\gkthing{14}}

{\newif\ifnotend
\notendtrue
\def\veclist{ABCDEFGHIJKLMNOPQRSTUVWXYZabcdefghijklmnopqrstuvwxyz.}
\def\top#1#2.{#1}
\def\tail#1#2.{#2.}
\loop\expandafter\xdef\csname v\expandafter\top\veclist\endcsname%
{{\noexpand\bf\expandafter\top\veclist}}
\edef\veclist{\expandafter\tail\veclist}
\if\veclist.\notendfalse\fi\ifnotend\repeat}
\def\cN{{\cal N}}
\def\d{{\rm d}}

\def\bolOm{\mbox{\boldmath$\Omega$}}
\def\vOmega{\bolOm}
\def\vPhi{\mbox{\boldmath$\Phi$}}

\def\Nbody{{\tt NBODY6}}

\def\e{\mathrm{e}}

\def\fracj#1#2{{\textstyle{#1\over#2}}}

\title[Relaxation of spherical stellar systems]
{Relaxation of spherical stellar systems}

\author[Jun Yan Lau \& James Binney]{
  Jun Yan Lau$^1$, James Binney$^1$\thanks{E-mail: binney@thphys.ox.ac.uk}\\  
  $^1$Rudolf Peierls Centre for Theoretical Physics, Clarendon Laboratory,
  Parks Road, Oxford, OX1 3PU, UK
}

\begin{document}
\maketitle

\begin{abstract}
10\,000 simulations of 1000-particle realisations of the same cluster are
computed by direct force summation. Over three crossing times the original
Poisson noise is amplified more than tenfold by self-gravity.
The cluster's fundamental dipole mode is strongly excited by Poisson noise,
and this mode makes a major contribution to driving diffusion of stars in
energy. The diffusive flow through action space is computed for the
simulations and compared with the predictions of both local-scattering
theory and the Balescu-Lenard (BL) equation. The predictions of local-scattering
theory are
qualitatively wrong because the latter neglects self-gravity. These results
imply that local-scattering theory is of little value. Future work on cluster
evolution should employ either N-body simulation or the BL
equation. However, significant code development will be required to make use
of the BL equation practicable. 
\end{abstract}


\section{Introduction} \label{sec:intro}

The central idea of stellar dynamics is that in a first approximation stars
move in the `mean-field' gravitational potential that one obtains by
smearing the mass of each star over a region that is somewhat larger than the
local inter-star distance. Then in a second step one considers how stars
drift between orbits in the mean-field potential on account of differences
between the actual potential of the cluster and the mean-field potential.
This process of drift between orbits is termed `dynamical relaxation' because
it drives secular increase in a cluster's entropy. Relaxation causes stars
with larger masses to be more strongly concentrated towards the cluster
centre than less massive stars (`equipartition') and the cluster's central
density to increase while its halo becomes more extended (`core collapse').
It also causes the mass of the cluster to diminish through stars being
occasionally accelerated to velocities larger than the local escape speed
(`evaporation'). 

These general principles have been understood since the seminal work of
\cite{Ed16}, \cite{Jeans1915} and \cite{Henon1961}. For decades it was not
feasible to follow the relaxation of a globular cluster by brute-force
computation. Hence relaxation was investigated, both analytically and
numerically, by adopting an ansatz for the difference between the
gravitational field of the cluster and its mean-field model. The ansatz was
that this consisted of the field of a point mass around each star, the idea
being that close to each star the mean field of the cluster is negligible
compared to the Kepler field, while far from each star the reverse condition
holds \citep{Chandra1949}. Hence relaxation can be modelled by summing large
numbers of deflections as pairs of stars scatter off each other at relatively
close quarters. Each such scattering could be computed via a hyperbolic
Kepler orbit of the reduced particle.

When this `local' relaxation theory is worked out in detail, the rate of
relaxation emerges as an integral over the impact parameters $b$ of these
hyperbolic orbits that diverges as $\ln b$ at large $b$ \citep[e.g.][]{GDII}.
This weak divergence was mastered by imposing a largest impact parameter to
be considered, $b_{\rm max}$. \cite{Chandra1949} took $b_{\rm max}$ to be the
local inter-particle distance. \cite{Theuns1996} argues that it should rather be
the larger of the cluster's core radius and distance to the cluster centre.
Since the divergence is only logarithmic, the relaxation rate computed using
either of these choices of $b_{\rm max}$ does not differ hugely, even though
the chosen values of $b_{\rm max}$ are typically very different.

Quite recently \cite{Heyvaerts2010} and \cite{Chavanis2012} developed a radically
different approach to relaxation. A rather formal analysis in the framework
of angle-action variables and using potential-density pairs to solve
Poisson's equation yields the `Balescu-Lenard equation', hereafter the BL
equation. This equation states
that a star diffuses through phase space as a consequence of interacting
resonantly with other stars in the cluster.  \cite{Chavanis2012} does not start
from the assumption that relaxation proceeds by interaction in pairs, but
arrives at this conclusion as a consequence of the structure of the
Boltzmann equation
\[
{\p f\over\p t}+[f,\Phi]=0.
\]
Here $f(\vx,\vv,t)$ is the one-particle distribution function (DF),
$\Phi(\vx,t)$ is the cluster's full (fluctuating) potential and $[,]$ denotes
a Poisson bracket. When $f$ and
$\Phi$ are decomposed into mean-field and fluctuating parts $f_0$, $f_1$,
$\Phi_0$ and $\Phi_1$, it follows that the evolution of $f_0$ (i.e.,
relaxation) is given by
\[
{\p f_0\over\p t}=\ex{[\Phi_1,f_1]}.
\]
That is, relaxation is driven by correlations between the fluctuating parts
of the potential and the DF. Two processes drive correlations: (i) by
Poisson's equation, the potential will be deeper where there are more
particles, and (ii) there will be more particles where the potential pushes
particles together. Mechanism (ii) operates even on a population of test
particles, whereas mechanism (i) operates only when particles bear mass.
Roughly, mechanism (ii) is responsible for evaporation and mechanism (i) is
responsible for equipartition and dynamical friction.

Thus the sophisticated BL equation confirms Chandrasekhar's
ansatz that relaxation occurs through interactions in pairs. However, it does
not confirm that these interactions are local. In fact, it replaces the local
criterion with a requirement that the stars can resonate in the sense that
there exist vectors $\vn$, $\vn'$ with integer components such that
$\vn\cdot\vOmega+\vn'\cdot\vOmega'=0$, where $\vOmega$ and $\vOmega'$ are
(two- or three-dimensional) vectors formed by the stars' fundamental orbital
frequencies.

Unlike Chandrasekhar and his followers, Heyvaerts and Chavanis included the
self-gravity of the cluster in their computation of the diffusion rate. Thus
stars interact with each other not through the vacuum but through the
polarisable medium formed by the rest of the system. \cite{JT1966}
first demonstrated that polarisation effects could be important by computing
the wake that a mass $M$ that is on a circular orbit in stellar disc raises
in the surrounding star field. The temperature of the disc is best quantified
by the $Q$ parameter \citep{To64}, which is the ratio of the velocity
dispersion in the disc to the critical value below which the disc is
(Jeans) unstable to axisymmetric disturbances. \cite{JT1966} showed
that for a realistic value $Q\simeq1.4$ the mass in the wake is several times
larger than $M$. Hence the effective mass of a disc star is several times
larger than its real mass because it polarises the disc around it, and we
must expect this increase in mass to facilitate exchanges of energy and
momentum with other disc stars. 

The BL equation tells us that  each star shakes the system at its natural
frequencies $\vn\cdot\vOmega$. If the system has a natural mode of
vibration near this frequency, the star excites this mode, and the mode's
energy  may be absorbed by a star that is quite distant but also happens
to resonate with the mode.

\cite{FouvryPMC2015} showed that inclusion of self gravity accelerates the
relaxation of a razor-thin disc by a factor nearer 1000 than 10 on account of
swing amplification, which \cite{JT1966} first worked out for a stellar
disc following its discovery in a gas disc by \cite{GoldreichDLB}.  Stars,
afforced by their wakes, launch running spiral waves, which are
swing-amplified at their corotation annulus and are finally resonantly
absorbed at a Lindblad resonance. Thus because disc stars communicate via
running waves that have an amplifier on the line, their interaction is by no
means local. The interaction is also $\sim1000$ times stronger than it would
be in the absence of self gravity. This strengthening, together with the
insight that the coupling is resonant, resolved a decades-old puzzle as to
why N-body simulations of discs that only have stable normal modes develop
strong spiral structure and then bars \citep{SellwoodC2014,FouvryPMC2015}.

Thus the BL equation has led to a major advance in our
understanding of stellar discs.  \citealt{Hamilton2018} (hereafter H18) asked
what its implications were for globular clusters, which unlike discs, have
been thought to be adequately described by  local-scattering theory. After
all, there is no known analogue of the swing amplifier in a globular cluster.

The BL equation describes the diffusion of stars through action
space, and provides a prescription for the computation of the relevant
diffusion coefficients.  Local theory provides an alternative prescription
for computing the diffusion coefficients, so key questions are (i) do the two
frameworks yield materially different coefficients? and, if so, (ii) which
framework's coefficients are more accurate?

A razor-thin disc and a spherical cluster both have an effectively
two-dimensional action space (H18).  To compute the BL
coefficients at a location $\vJ$ in this space, one must (i) identify lines
in action space on which the resonance condition
$\vn\cdot\vOmega+\vn'\cdot\vOmega'=0$ is satisfied, and then (ii) integrate a
certain function along each such line (i.e., vary $\vJ'$ so the condition
remains satisfied). In the case of a razor-thin disc, it is physically
plausible that a small number of lines dominate the diffusion coefficients,
but in a spherical cluster this is not true. Given that it was not feasible
to evaluate the necessary integrals along all resonant lines, H18
evaluated them along lines defined by integers $|n_i|\le2$. Since this scheme
led them to omit very many resonant lines, including the lines most likely to
be effectively included by local theory, it would be natural for their values
of the diffusion coefficients to be smaller than the coefficients computed in
the local approximation. But their values were {\it not} smaller. By
repeating their calculation with self-gravity neglected, they were able to
show that their diffusion coefficients are strongly enhanced by self-gravity,
which is neglected in the local approximation. This finding cast doubt of the
applicability of the local approximation to globular clusters, but in light
of the incompleteness of their treatment H18 had no basis to
claim that their diffusion coefficients were more accurate than those yielded
by the local approximation.

In a spherical system, the second component of the vector $\vn$ in a resonant
condition is always the angular-momentum quantum number $\ell$. Hence, H18's
sums over $n_i\le2$ were sums over certain monopole, dipole and quadrupole
distortions of the cluster. That is, they computed the contributions to
diffusion coefficients that arise from stars exciting/damping simple global
distortions of the cluster. In these circumstances it is no surprise that
neglect of self-gravity much diminishes the diffusion coefficients. For
example, the simplest  dipole ($\ell=1$) mode involves displacing the core
of the cluster with respect to its envelope. If this is done with
self-gravity neglected, so in a fixed potential, the core being anchored by
the potential, can barely move.  By contrast, when self-gravity is included
the core is free to move because it will take its potential with it. Similar,
though weaker, arguments apply to motion of the envelope.

H18 showed that their (incomplete) BL coefficients and the
coefficients from the local approximation predict qualitatively different
diffusive flux vectors in action space. So the two theories will predict
different evolutionary tracks.  Given that the work of H18 strongly suggests
that the cluster's low-order global modes, which are excluded from the local
approximation, contribute significantly to the diffusion coefficients, one
can have no confidence in the essential validity of  the local approximation.
Yet  decades of work by many groups has been underpinned by the local
approximation. Can it really be seriously in error?

It is now feasible to integrate directly the equations of motion of clusters
with realistic numbers of stars. By examining orbital diffusion in such
simulations, one can rigorously test any theory of diffusion.  Surprisingly
few studies have done this, however. An early attempt was that of
\cite{Theuns1996}, who measured the second-order diffusion coefficient $\d(\Delta
E)^2/\d t$ in the energies $E$ of stars in \cite{King1966} models with
different particle numbers, and compared these measurements with values
obtained in the local approximation.  He concluded that `overall' the values
agreed to an `impressive' extent although the measured values were larger
than predicted by factors 1.5--2 in a more centrally concentrated model and
smaller than expected by similar factors in a less centrally concentrated
model. \cite{Kim2008} used a Fokker-Planck (FP) code, which is based on the
local approximation to compute the evolution of rotating King models with
$10\,240$ stars and compared its predictions for the evolution of the central
density, velocity dispersion, rotation rate and velocity anisotropy with what
they found by direct integration of $N$ bodies. The FP predictions were
qualitatively correct but gave rise to quantitative discrepancies.

Given that the local approximation has in the Coulomb logarithm an
effectively free parameter, it calls for more rigorous validation than
checking consistency with a few numbers.  A more demanding  comparison of the
predictions of local theory with direct N-body integration is by comparing
diffusive fluxes at each point in action space. It is these fluxes that drive
the evolution in density and velocity profiles studied by \cite{Kim2008}, and
diffusion coefficients in actions $\vJ$ are more meaningful than the
diffusion coefficients in energy studied by \cite{Theuns1996} because changes
in $E$, unlike changes in the adiabatic invariants $\vJ$ are not solely
caused by the fluctuating component of the potential, but have a significant
contribution from the secular evolution of the mean-field potential $\Phi_0$.

In this paper we use direct N-body integration to determine the diffusive
flux $\vF(\vJ)$ in action space for a cluster of $1000$ stars that has the
isochrone DF \citep{Henon1960}, and we compare this $\vF(\vJ)$ with the
fluxes computed by H18 for the same system from (i) the local approximation,
and (ii) the BL equation. Section~\ref{sec:simuls} describes the simulations,
Section~\ref{sec:centre} discusses the key issue of choice of cluster centre
and presents evidence for powerful excitation of a dipole distortion.
Section~\ref{sec:Phi} characterises the fluctuating component of a cluster's
gravitational field and links this to diffusion of stars in energy.
Section~\ref{sec:drift} presents the drift of stars through action space and
discusses rates of entropy generation. Section~\ref{sec:conclude} sums up.
Appendices summarise units for N-body models and the computation of diffusion
coefficients for energy in the local-scattering approximation.

\section{The simulations}\label{sec:simuls}

The isochrone is a spherical system of mass $M$ that has the gravitational potential
\[
\Phi(r)=-{GM\over b+\sqrt{b^2+r^2}},
\]
where $b$ is the system's scale radius.
We created $10\,000$ realisations of the isochrone, each with  $1000$
equal-mass stars. Each star's  radial coordinate $r$ was chosen by picking a
random number $\xi_1$ uniformly on $(0,1)$ and then inverting
the isochrone's cumulative mass distribution
\[
\xi_1(r)={1\over M}\int_0^r\d r\,r^2\rho_{\rm I}(r),
\]
 where $\rho_{\rm I}(r)$ is given by equation (2.49) of \cite{GDII}. 
Once $r$ had been chosen, a second  random number $\xi_2$ was picked and the
particle's speed $v$ was determined by inverting
\[
\xi_2(v)={\int_0^v\d v\,v^2f_{\rm I}[\fracj12v^2+\Phi(r)]\over
\int_0^\infty\d v\,v^2f_{\rm I}[\fracj12v^2+\Phi(r)]},
\]
where $f_{\rm I}(E)$ is the DF of the isochrone \citep[][eqn 4.54]{GDII} and
the function of $v$ on the right of this equation was inferred by
interpolation on a grid of numerically computed values.  The directions of
the position and velocity vectors $\vr$ and $\vv$ were chosen randomly. This
sampling procedure was validated by binning the stars in energy and comparing
the resulting histogram with the analytically determined differential energy
distribution \citep[][\S4.3.1(b)]{GDII}. 

\Nbody\ \citep{Aarseth2000} was then used to integrate the equations of motion of
each realisation. \Nbody\ automatically boosts to the zero-momentum frame of
the initial conditions, and rescales positions and velocities such that
distances and times are measured in units of a length $r_N$ and time $t_N$
that are derived from the potential and kinetic energies of the initial
conditions (Appendix \ref{app:units}). The ratio $r_N/b$ varies slightly from
realisation to realisation, but in Appendix \ref{app:units} we show that
$b=0.2375\ex{r_N}$. The isochrone has a natural timescale
\[
t_{\rm I}\equiv(b^3/GM)^{1/2}.
\]
In Appendix~\ref{app:units} we show that the N-body time unit $\ex{t_N}=8.642t_{\rm
I}$. In the following  distances and times are given in units of $b$ and
$t_{\rm I}$,
ignoring the small scatter in $r_N$ and $t_N$ between realisations.

The equations of motion of each
realisation were integrated for a time $10t_N$, with snapshots of the
phase-space coordinates saved after each elapse of $\delta t=0.5t_N$. Thus
the data comprise 10\,000 sets of 21 1000-particle snapshots.

\section{Choice of centre}\label{sec:centre}

We wish to study relaxation via changes in quantities that in the mean-field
model are constants of motion. To do this we have to choose a centre. Three
possible choices spring to mind:

\begin{itemize}
\item[1.] The mean-field centre, i.e., the point $r=0$ around which each
realisation was assembled. By symmetry it must coincide with the expectation
value of any credible estimator of the centre. The mean-field centre is the
only centre that is defined in the context of the BL equation.

\item[2.] The barycentre. Momentum conservation ensures that a realisation's
barycentre defines an inertial frame of reference. Moreover, in our chosen
zero-momentum frame the barycentre is absolutely stationary. A severe
disadvantage of the barycentre is that it is sensitive to the small number of
stars that are at large radii, and the distribution of its offsets from the
mean-field centre is long-tailed: a Gaussian fit to any component of this
offset yields a dispersion $s_{\rm b}\sim2.3b$ yet  $\sim34$ percent of
realisations had barycentres that lay more than $10b$ from the
mean-field centre. In view of this finding, we did not use the barycentre.

\item[3.] The potential centre. Our procedure for initialising the
simulations makes it exceedingly unlikely that a realisation initially
includes a hard binary, and the timescale for hard binaries to form via
three-body interactions  is much longer than the
duration of our integrations \citep[][\S7.5.7(c)]{GDII}. So binaries are not a concern. This being so,
the star with the most negative potential energy is likely to lie close to
a cluster's physical centre. Consequently, serious consideration should be
given to computing actions with respect to this centre.  

\end{itemize}

The problem with the use of the potential centre is its erratic
motion. In as much as it is displaced from the barycentre, it must be in
orbit around that fixed point. This orbit can be considered to be a
consequence of the force that the somewhat
lop-sided halo exerts on the core. On top of this smooth motion, the potential
centre occasionally jumps discontinuously when the distinction of being the most bound
star switches between stars. These jumps can be made more frequent but
smaller in magnitude by adopting as the potential centre not the location of
the most bound particle but the barycentre of the $n$ most bound particles.
We found $n=30$ to be a sensible choice.

If we use the mean-field centre to compute actions, those actions will have
non-zero rates of change because the actual potential  $\Phi$ differs from
the mean-field potential $\Phi_{\rm mf}$. For this centre the rate of change of any action
is 
\[\label{eq:Jdot}
{\d J\over\d t}=[J,\Delta\Phi]
\]
where $[,]$ is a Poisson bracket and
\[\label{eq:PhiMF}
\Delta\Phi\equiv\Phi-\Phi_{\rm mf}.
\]
 If we define actions relative to  the potential centre,
the actual potential $\Phi$ will be much more nearly a function of distance
fronm the potential centre, so if
equation (\ref{eq:Jdot}) applied with $\Delta\Phi$ equal to the 
difference between $\Phi$ and the closest function of distance from the
potential centre, $|\dot J|$ would be
small. However, since the potential centre does not define an inertial
coordinate system, terms describing pseudo-forces would need to be added to the right
side of equation (\ref{eq:Jdot}), and typical values of $|\dot J|$ would
probably be comparable to the values encountered when using the mean-field
centre. 

In the following actions will be defined with respect to the
mean-field centre.

\subsection{Dynamics of the potential centre}

The practical difficulties with defining integrals of motion with respect to
the potential centre do not alter the fact that it has physical significance.
In fact, understanding its motion is crucial for understanding how clusters
relax. 

\begin{figure}
\includegraphics[width = \columnwidth]{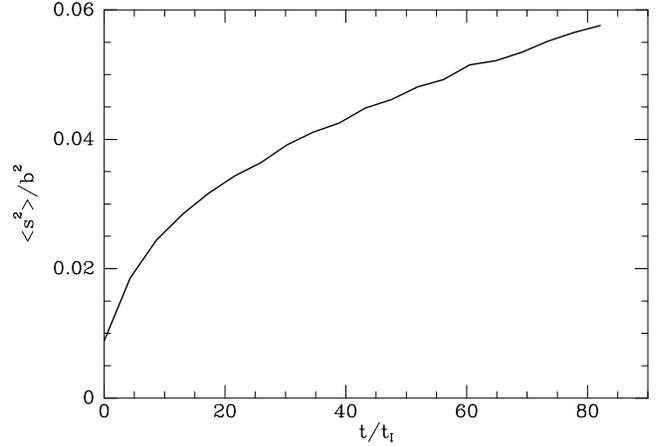}
\caption{The dispersion of the distances $s$ between the potential and
mean-field centres of realisations as a function of the time for for which
\Nbody\ has advanced the phase-space coordinates.}\label{fig:Maxwell}
\end{figure}

Given that the
potential centre of each cluster is independently and isotropically scattered
around the mean-field centre, by the central limit theorem the distribution
of the distances $s$ between these centres must be Maxwellian.
Fig.~\ref{fig:Maxwell} shows  the variance $\ex{s^2}$ of these distances as a
function of integration time $t$. The variance starts  small and grows as
\Nbody\ advances the particles. We confirmed that it does {\it not} grow if the
mean-field potential is used to advance particles. Thus the growth of
$\ex{s^2}$ evident in Fig.~\ref{fig:Maxwell} is an effect of self-gravity.

The initial variance is small because it
is entirely due to Poisson noise in sampling the core of the cluster. As the
equations of motion are integrated, it grows at a rate that for $t\ga t_N$
settles to a constant. Thus the variance tends to linear dependence on
integration time: $\ex{s^2}\propto t$. This is the signature of a
random walk rather than steady motion, which would imply $s\propto t$. This
establishes that the potential centre experiences significant acceleration.

Given that each cluster's barycentre is absolutely fixed, the potential
centre cannot execute an unconstrained random walk, for that would eventually
take it arbitrarily far from the barycentre. Hence the potential centre must
executes a tethered random walk, like a randomly kicked ball that is attached
by a piece of elastic to a fixed peg. The standard deviation of the distances
$s_{\rm b}\sim2.3b$ of the barycentres from the mean-field centre sets the scale of
the region over which we expect the potential centre to wander.  Since this
scale is bigger than the standard deviations $\ex{s^2}^{1/2}$ plotted in
Fig.~\ref{fig:Maxwell}, the elastic is slack throughout our integrations
because they start with the potential centre anomalously close to the
mean-field centre. The elastic being slack, the potential centre is for the
moment executing an essentially random walk and $\ex{s^2}\propto t$ in
consequence.

\begin{figure}
\includegraphics[width = \columnwidth]{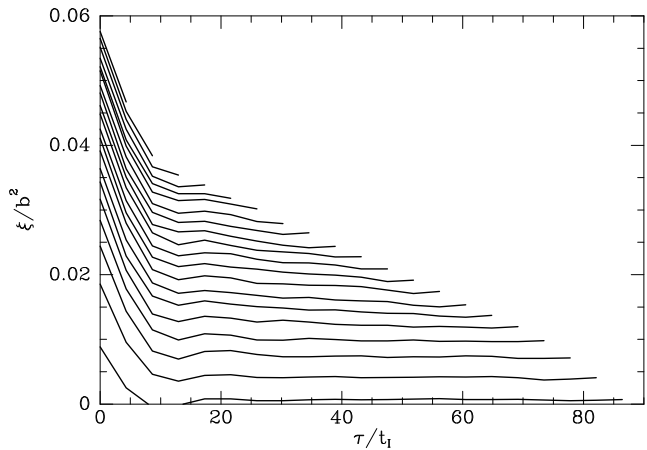}
\caption{The autocorrelations of the location of the potential centre
(eqn.~\ref{eq:auto}). The bottom curve uses all the data, while the next curve
up discards  the initial condition, and the curve above that discards the
first two snapshots, and so on up the sequence of curves.}\label{fig:auto}
\end{figure}

Fig.~\ref{fig:auto} indicates that the picture just deduced is an
oversimplification. It shows estimates of the autocorrelation
\[\label{eq:auto}
\xi(\tau)={1\over N}\sum_{i=1}^N{1\over T-\tau}\int_0^{T-\tau}\!\!\d
t\,\,\vx(t)\cdot\vx(t+\tau)
\]
of the positions of the potential centres. The bottom curve uses all
snapshots, while the next curve up discards the first snapshot but uses all
the other ones, and the second curve up discards the first two snapshots, and
so on. The more snapshots are discarded, the shorter is the range of delays
$\tau$ that can be probed. Aside from this progressive shortening of the
curves, they have remarkably similar shapes: a steep
initial decline is followed by a small bump before a plateau is reached. Note
that this structure in the bottom curve emerges from snapshots that have no
overlap with the snapshots that give rise to the structure in the topmost
curves. Thus the structure is highly reproducible.

The curves' vertical offsets are a straightforward consequence of the steady
rise in $\ex{s^2}$ shown in Fig.~\ref{fig:Maxwell}: $\xi(0)$ is simply the
mean of $s^2$ over the snapshots used to compute $\xi$.

The horizontal plateaus of the curves are simple consequences of the random
walk we inferred above. In this picture $\vx(t+\tau)$ is
the sum of $\vx(t)$ and a series of random increments $\vdelta_i$, so
\[\label{eq:steps}
\xi(\tau)=\ex{\vx(t)\cdot\biggl[\vx(t)+\sum_i\vdelta_i\biggr]}.
\]
So long as the steps $\vdelta_i$ are distributed isotropically, the
expectation value of $\vx(t)\cdot\vdelta_i=0$, so $\xi(\tau)=|\vx(t)|^2$ is
independent of $\tau$.

\begin{figure}
\centerline{\includegraphics[width=.6\hsize]{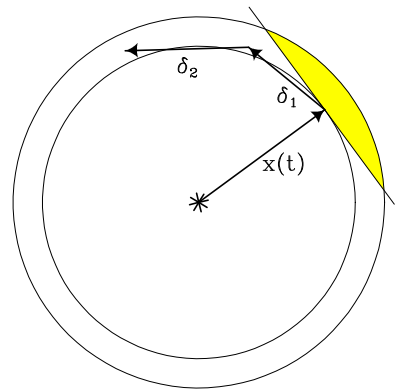}}
\caption{Schematic of the motion of the potential centre. The central star
marks the location of the mean-field centre and the three arrow heads mark
the locations of the potential centre in successive snapshots. The
projections of the two shift vectors $\vdelta_i$ on $\vx(t)$ are negative
because the potential centre is roughly in orbit around the mean-field
centre. The projection of $\vdelta_1$  is positive only if it leads to a
point in the small fraction of the annulus that is shaded yellow.}\label{fig:cartoon}
\end{figure}

What then is the cause of the initial steep drop in $\xi(\tau)$? Since this
feature occurs in curves that discard the initial conditions and the
immediately following snapshots, this steep decline cannot be connected to
our starting the simulations from anomalously regular initial conditions.
The argument just given to explain the plateaus, suggests that $\xi$ declines
initially because the first steps $\vdelta_i$ satisfy
$\vx(t)\cdot\vdelta_i<0$. This condition will be satisfied if steps have a
tangential bias. Fig.~\ref{fig:cartoon} illustrates this idea. If the motion
of the potential centre over short timescales resembles an orbit,
$\vx(t)\cdot\vdelta_i$ will be negative for of order half an orbital period.
Later it will be positive for a similar time. When this type of variation is
averaged over $10\,000$ realisations, in each of which the potential centre is
on a slightly different orbit, the positive and negative contributions to
$\xi(\tau)$ tend to cancel once  $\tau$ is comparable to an orbital period,
and a plateau in $\xi$ ensues.

This discussion leads to the tentative conclusion that the delay
$\tau\simeq10t_{\rm I}$ at which the plateaus start in Fig.~\ref{fig:auto} is the
time over which the motion of the potential centre resembles an orbit, and
after which it is dominated by noise. The noise is initially highly non-stationary
because the simulations start from anomalously regular initial conditions
that reflect Poisson noise in the absence of self gravity. As the equations
of motion are integrated, self-gravity amplifies this noise by a significant
factor.

\section{Global potential fluctuations}\label{sec:Phi}

We now quantify the stochastic perturbing potential $\Delta\Phi$, which is
the difference between the true and mean-field potentials.
\Nbody\ returns the value of the true potential at the locations of each
particle, so it is trivial to compute the value of $\Delta\Phi$ at each
particle's location. However, to make contact with the work of H18 we want to
decompose $\Delta\Phi(\vx)$ in spherical harmonics around the mean-field
centre. Unfortunately, there is no straightforward way to compute this
decomposition from the \Nbody\ values on account of the biased
distribution of particle locations. Consequently, we obtain the desired
expansion by using spherical harmonics to solve Poisson's equation afresh.
That is we compute the true potential from the standard formula \citep{GDII}
\[\label{eq:PhiY}
\Phi(r,\theta,\phi)=\sum_{\ell m}\Phi_{\ell m}(r)Y_l^m(\theta,\phi),
\]
where
\begin{align}\label{eq:multip1}
\Phi_{\ell m}(r) =  -\frac{4\pi G}{2l+1} \bigg[&\frac{1}{r^{l+1}} \int_0^r\d a\,
\rho_{\ell m}(a) a^{l+2}\cr
&\qquad + r^l \int_r^\infty\d a\, a^{1-l}\rho_{\ell m}(a) \bigg].
\end{align}
Here $(r,\theta,\phi)$ are polar coordinates around the mean-field centre and
\[\label{eq:multip2}
\rho_{\ell m}(a)\equiv\int\d^2\Omega\,Y_l^{m*}(\theta,\phi)\rho(a,\theta,\phi).
\]
 Our knowledge of the density  $\rho(\vx)$ is limited to the Monte Carlo
sample   provided by \Nbody. Given this fact, a natural procesdure is
to estimate $\rho(\vx)$ by binning particles in spherical shells of finite
thickness and then discretising the integrals in equation (\ref{eq:multip1})
into sums over bins. It is not hard to show that the statistical noise in the
resulting estimate of $\Phi_{\ell m}$ is essentially invariant as we reduce the
thickness of shells: the increase in the noise level of a each $\rho_{\ell m}$ is
precisely compensated by the increase in the number of shells that contribute
effectively to $\Phi_{\ell m}(r)$ at a given radius. Hence, it is safe to reduce
the width of shells so each shell contains at most one particle. In this
limit the sum over shells in equation (\ref{eq:multip1}) becomes a sum over
particles, with $\rho_{\ell m}(a)$ replaced by the value of $Y_l^{m*}$ at the
particle's angular location.

\begin{figure}
\includegraphics[width=\hsize]{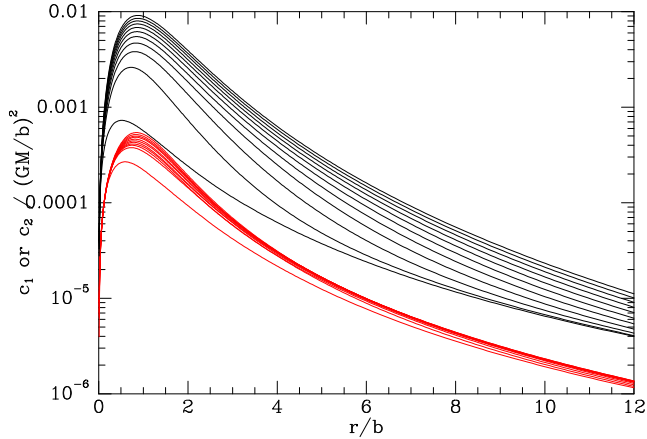} \caption{The average power
of the dipole (black) and quadrupole (red) components of the potential as
functions of radius at a series of times during which \Nbody\ has integrated
the equations of motion.}\label{fig:dipole_mag}
\end{figure}

\begin{figure}
\includegraphics[width=\hsize]{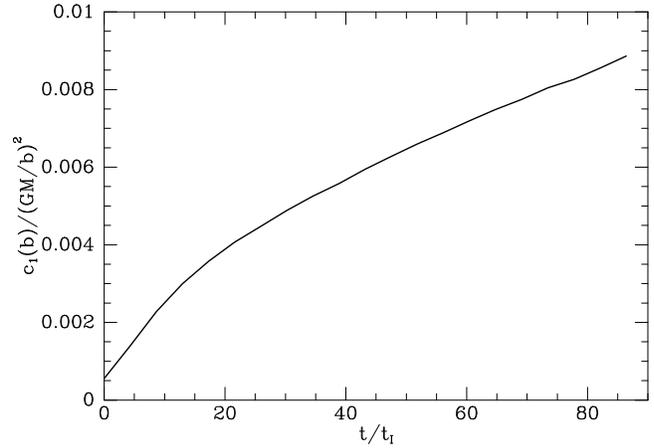}
\caption{The average dipole power $c_1(b)$ at $r=b$ as a function of
time.}\label{fig:dipole1}
\end{figure}

Each black curve in Fig.~\ref{fig:dipole_mag} shows the average over all
realisations of the spectral power
\[\label{eq:c1}
c_1(r)={\sum_{m=-1}^1\big|\Phi_{1m}(r)\big|^2}
\]
of the dipole component of the potential at a given time. The lowest curve is
for $t=0$, i.e., the initial conditions, and as the time of integration
increases, the corresponding curve moves upwards.  Fig.~\ref{fig:dipole1}
shows $c_1(b)$ as a function of
time. The curve strongly resembles the curve of $\ex{s^2}$ shown in
Fig.~\ref{fig:Maxwell}. In particular, at later times $c_1\propto t$.  The
growth of the dipole in $\Phi$ by a factor in excess of 10 puts a lower limit
on the extent to which the initial conditions are unnaturally symmetric
because they neglect the long-range correlations that grow rapidly when stars
move in the self-consistent potential.

Each  red curve in Fig.~\ref{fig:dipole_mag} plots the ensemble average of
the magnitude
 \[\label{eq:c2}
c_2(r)={\sum_{m=-2}^2\big|\Phi_{2m}(r)\big|^2}
\]
 of the quadrupole
component of $\Phi(\vx)$. The curves are similar to the dipole curves, but
depressed by a factor of order 10 and displaying weaker growth of quadrupole
distortions as \Nbody\
advances the particles.

\begin{figure}
\includegraphics[width=\hsize]{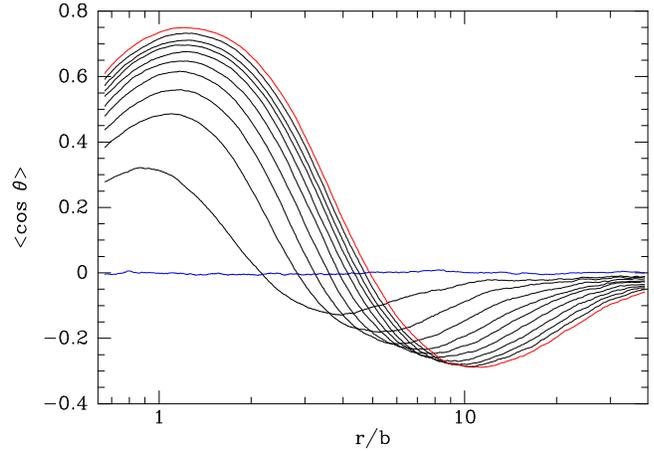}
\caption{Each curve shows for a given time the average over realisations of
$\cos\theta$, where $\theta$ is the angle between the dipole formed by the
particles within radius $r$ of the mean-field centre and the dipole formed by the
remaining particles. The blue and red curves are for the initial conditions
and the latest time, respectively.}\label{fig:angle}
\end{figure}

From the perspective of perturbation theory, displacement of the physical
centre from the mean-field centre causes the perturbing potential
$\Delta\Phi$ (eqn~\ref{eq:PhiMF}) to be dominated by a dipole term, so the
results of this section are natural consequences of the random walk we
inferred in the last section.

The dominance of the dipole component of $\Phi$ explains why
H18 found from the Balescu-Lenard equation that the rate of relaxation
was dominated by the dipole of the perturbing potential. The much reduced
values of $\ex{s^2}$ when the mean-field potential is used to advance
particles corresponds to strong reduction in the relaxation rate when
H18 switched off self gravity. On account of the random nature of
the motion of the physical centre, the perturbing dipole potential has the
stochastic nature assumed in the derivation of the Balecu-Lenard equation.

H18 found that the largest contribution to the relaxation rate came from the
$\ell=1$ mode in which the cluster's core and halo move in antiphase. In such
a mode, the dipoles in the density distribution at small and large $r$ should
have opposite directions.  The complex numbers $\rho_{1m}$ define a vector
\[\label{eq:D}
\vD\equiv\left(\surd2\Re\e\rho_{11},\surd2\Im{\rm m}\rho_{11},\rho_{10}\right).
\]
Each curve in Fig.~\ref{fig:angle} plots the average over all
realisations of
\[
\cos(\theta)\equiv{\vD_r^-\cdot\vD_r^+\over
|\vD_r^-||\vD_r^+|},
\]
where $\vD_r^-$ is the vector (\ref{eq:D}) defined by the dipole formed by
all the particles interior to radius $r$, while $\vD_r^+$ is the vector
associated with the remaining particles. The blue curve shows that in the
initial conditions the interior and exterior dipoles have random relative
orientations, regardless of the radius taken to define `interior'. As \Nbody\
advances the particles, anti-alignment of the dipoles emerges rapidly. We
take the radius at which a curve passes its minimum as the measure of the
boundary between the zones of opposite alignment: if $r$ is decreased from
this value, $|\ex{\cos\theta}|$ decreases because shells with interior
alignments are classified as exterior.  Conversely, if $r$ is increased
$|\ex{\cos\theta}|$ decreases because shells with exterior alignment are
classified as interior. Fig.~\ref{fig:angle} shows that in the first
snapshot, the boundary lies at $r\simeq4b$ while in the last snapshot the
boundary has reached $r\simeq11b$. In the initial snapshot
$|\ex{\cos\theta}|$ is very small for $r\ga10b$, indicating that shells
outside $10b$ have randomly aligned dipoles. As \Nbody\ advances the cluster,
the point at which $|\ex{\cos\theta}|$ becomes small moves rapidly outwards
as larger and larger shells align their dipole in opposition to the cluster's
core.

\subsection{Impact of global fluctuations}

How much relaxation can be attributed to the large-scale fluctuations in
$\Phi$ that we have just quantified? 

To answer this question we 
start from the equation for the rate of change of a particle's energy
$E=\fracj12 v^2+\Phi$:
\[
{\d E\over\d t}={\p\Phi\over\p t},
\] 
which yields the change $\Delta E$ in $E$ during an interval $\tau$ as
\[
\Delta E(\tau)=\int_{\rm orbit}^\tau\d t\,{\p\Phi\over\p t}.
\]
Squaring both sides and taking the expectation value over orbits we find
\begin{align}\label{eq:cPhi}
\ex{(\Delta E)^2}_\tau
&=\int_0^\tau \d t\int_0^\tau\d t'\,\ex{{\p\Phi\over\p t}{\p\Phi\over\p t'}}\cr
\end{align}
With the expansion of equation (\ref{eq:PhiY}) this becomes
\begin{align}\label{eq:YPhi}
\ex{(\Delta E)^2}_\tau
&=\sum_{\ell m\ell'm'}\int_0^\tau \d t\int_0^\tau\d t'\,\cr
&\times\Bigl\langle{\p\Phi_{\ell m}^*\over\p
t}{\p\Phi_{\ell'm'}\over\p t'}
Y_l^{m*}(\theta,\phi)Y_{\ell'}^{m'}(\theta',\phi')\Bigr\rangle.
\end{align}
Since we are taking the expectation over spherical clusters, we are
effectively integrating over angles. If the primed angles were independent of
the unprimed angles, the expectation value would vanish for $l>0$. But when
$t\simeq t'$, the primed and unprimed angles are similar, and the
expectation value approximates
\[\label{eq:Yorthog}
\ex{Y_l^{m*}(\theta,\phi)Y_{\ell'}^{m'}(\theta,\phi)}=(4\pi)^{-1}\delta_{l\ell'}\delta_{mm'}.
\]
Let $2\tau_\ell$ be the time during which the primed and unprimed angles are
sufficiently nearly equal for $(4\pi)^{-1}\delta_{ll'}\delta_{mm'}$ to be a
useful approximation to the expectation of the product of spherical harmonics
in equation (\ref{eq:YPhi}). Then we have
\[\label{eq:Ediff}
\ex{(\Delta E)^2}_\tau\simeq{1\over4\pi}\int_0^\tau \d
t\int_{t-\tau_\ell}^{t+\tau_\ell}\!\!\!\d t'\,\sum_{\ell m} 
\Bigl\langle{\p\Phi_{\ell m}^*\over\p t}{\p\Phi_{\ell m}\over\p t'}\Bigr\rangle.
\]
If $\Phi$ were a stationary random process, the integrand of the integral
over $t$ would be independent of $t$ and equation (\ref{eq:Ediff}) would
predict that $\ex{(\Delta E)^2}_\tau\propto\tau$ as we expect from a
diffusive process. In our simulations, $\Phi$ is not a stationary process
because the simulations start from anomalously small $\Delta\Phi$, but
$4\pi\d\ex{(\Delta E)^2}_\tau/\d\tau$ is still given by the $t$ integrand of
equation (\ref{eq:Ediff}).

\begin{figure}
\includegraphics[width=\hsize]{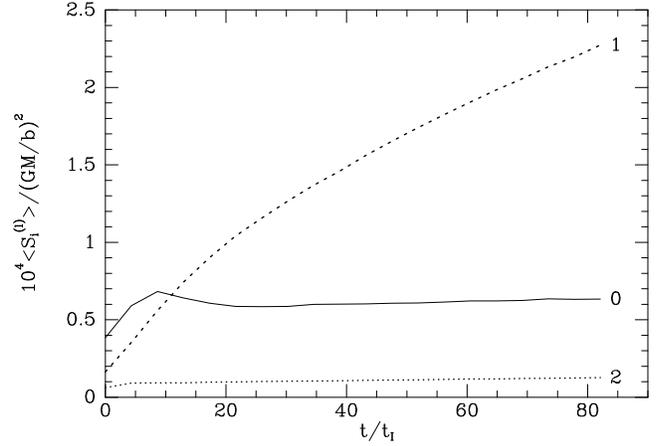} \caption{Estimates of the the
mean-square fluctuations in the monopole (full), dipole (dashed) and
quadrupole (dotted) components of the potential (eqn.~\ref{eq:defSell})
versus simulation time.}\label{fig:S}
\end{figure}

\begin{figure}
\includegraphics[width=\hsize]{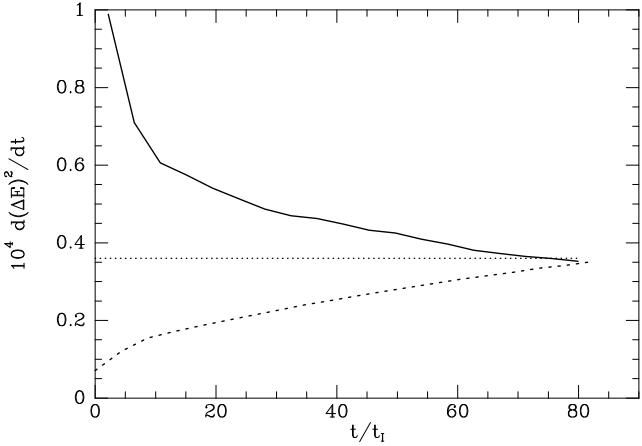} \caption{
Three estimates of $\d\ex{(\Delta E)^2}/\d t$. The full curve is
obtained by differencing  the energies of particles between snapshots of
simulations. The dotted curve is the analytic prediction (\ref{eq:steady}) of local scattering
theory. The dashed curve is obtained from equations (\ref{eq:Ediff2}) and
(\ref{eq:StoPhi}) by measuring
the amplitudes of the fluctuations in the monopole, dipole and quadrupole
contributions to the potentials of individual simulations and taking the
characteristic duration of a fluctuation to be $\overline{\tau}=8.6t_{\rm I}$.}\label{fig:E} 
\end{figure}

To estimate the typical size of $\p\Phi_{\ell m}/\p t$ we first note the
expression for the fluctuating part of the potential:
\[
\delta\Phi(\vx,t)=\sum_{\ell m}\Phi_{\ell m}(|\vx|,t)Y_\ell^m(\vx).
\]
This yields the mean-square amplitude of the fluctuations as
\[
\ex{\delta\Phi^2(r)}\simeq{1\over4\pi}\sum_{\ell m}\ex{|\Phi_{\ell m}(r)|^2}.
\]
Hence in  equation (\ref{eq:Ediff}) we may use the estimate
\[
{1\over4\pi}\sum_{\ell m} 
\Bigl\langle{\p\Phi_{\ell m}^*\over\p t}{\p\Phi_{\ell m}\over\p t'}\Bigr\rangle
\simeq{\ex{(\delta\Phi)^2}\over\overline{\tau}^2},
\]
 where $\overline{\tau}$ is an appropriate average of the $\tau_\ell$. With
 this estimate equation (\ref{eq:Ediff}) becomes
\[\label{eq:Ediff2}
\ex{(\Delta E)^2}_\tau\simeq\int_0^\tau \d
t\int_{t-\overline{\tau}}^{t+\overline{\tau}}\!\!\!\d t'\, 
{\ex{(\delta\Phi)^2}\over\overline{\tau}^2}
\simeq\int_0^\tau\!\!\d t\,{\ex{(\delta\Phi)^2}\over\overline{\tau}},
\]
where the second equality requires that the expectation now involves
averaging over particle radii as well as angles.

Now we estimate $\ex{(\delta\Phi)^2}$ by ranking stars by radius in each cluster
snapshot, and for the particle ranked $i$ in the $\alpha$th realisation we
compute the values for $\ell=0,1,2$ of
\[
\Phi^{(\ell)}_{\alpha i}\equiv\sum_m\Phi_{\ell m}(|\vx_{\alpha i}|)\,Y_l^m(\vx_{\alpha i}),
\]
i.e., the contributions of the monopole, dipole and quadrupole to the
potential energy of the $i$th ranked particle in a particular realisation. 
Then we compute  the sums over realisations
\[\label{eq:defSell}
S^{(\ell)}_i\equiv{1\over\cN}\sum_{\alpha=1}^\cN
\ex{\bigl(\Phi_{\alpha i}^{(\ell)}-\Phi_{\rm tot}^{(\ell)}(\vx_{\alpha i})\bigr)^2},
\]
where $\Phi_{\rm tot}^{(\ell)}$ is obtained from the potential computed from
the locations of all 10 million stars at the given time. When we average in
this way over all realisations we obtain a potential that has negligible
dipole and quadrupole contributions, but a monopole contribution that is very
similar to, but significantly different from the analytic isochrone
potential. $S_i^{(\ell)}$ is a measure of the amplitude of the fluctuations
in $\Phi$ contributed by the $\ell$ multipole at the typical radius of the
$i$th ranked particle. Hence the quantity $\ex{(\delta\Phi)^2}$ encountered
above is just $S_i^{(\ell)}$ summed over $\ell$ and averaged over $i$:
\[\label{eq:StoPhi}
\ex{(\delta\Phi)^2}=\sum_{\ell}\overline{S}^{(\ell)}
\hbox{ where }
\overline{S}^{(\ell)}\equiv{1\over N}\sum_i S_i^{(\ell)}.
\]
The full, dashed and dotted curves in
Fig.~\ref{fig:S} show for $\ell=0,1,2$, respectively, the  values of
$\overline{S}^{(\ell)}$ versus time. Initially, the
monopole is largest, but it is soon overtaken by the dipole, which continues
to rise throughout the simulated period, whereas the monopole stagnates. The
quadrupole continues to rise but is much smaller than either the monopole or
dipole. 

The dashed curve in Fig.~\ref{fig:E} shows the
resulting estimate from equations (\ref{eq:Ediff2}) and (\ref{eq:StoPhi}) 
of the rate of growth of $\ex{(\Delta
E)^2}$ when one sets $\overline{\tau}=t_N=8.6t_{\rm I}$
(Appendix~\ref{app:units}). The  full curve in Fig.~\ref{fig:E} shows the
actual rate of growth of $(\Delta E)^2$. Although the shapes of the two
curves are very different, at the latest time they agree in magnitude better
than one has any right to expect given the approximations involved in the
computation of the dashed curve.


In  Appendix~\ref{app:DE} we quote the formulae for the diffusion
coefficients in energy that are yielded by local scattering
theory. When the second-order diffusion coefficient is integrated over the
isochrone, we find
\[
\overline{D}_2\equiv
\int\d E\, g(E)f(E)\ex{{\d E^2\over\d t}}=1.20\ln\Lambda\times10^{-5}{\sigma^5\over b},
\]
 where $\sigma^2\equiv GM/b$.  The theory predicts that when we average
$\d\ex{\Delta E)^2}/\d t$ over all particles, we should find
that
\[\label{eq:steady}
{\d\ex{(\Delta E)^2}\over\d t}=\overline{D}_2.
\]
The dotted line in Fig.~\ref{fig:E} shows this
prediction when we adopt $\ln\Lambda=3$. 

Interpretation of Fig.~\ref{fig:E} is not straightforward. With our chosen
value of $\ln\Lambda$, local scattering theory yields the correct order of
magnitude for $\d\ex{(\Delta E)^2}/\d t$ but it predicts that the latter
should be constant rather than falling. The hypothesis that relaxation is
driven by large-scale fluctuations also yields the right order of magnitude
for $\d\ex{(\Delta E)^2}/\d t$ and it correctly predicts that the latter's
value should be a strong function of time rather than constant.
Fig.~\ref{fig:E}, however, suggests that it should be a rising rather than a
falling function of time.

Careful consideration reveals that this prediction is likely be an artifact
of how we have estimated the right side of equation (\ref{eq:Ediff}): we have
taken the time derivatives of $\Phi_{\ell m}$ to be their values divided by
the characteristic time $t_N$. This procedure makes sense once the system has
settled to an approximate steady state, but initially it is misleading
because the fluctuations start from anomalously small amplitudes, so we
estimate that they initially have very small time derivatives.
Fig.~\ref{fig:dipole1} shows clearly that the dipole's time derivative is
initially large, presumably because over-dense regions are falling together,
even though its initial magnitude is small. A better estimate of the initial
time derivatives of the $\Phi_{\ell m}$ would be the values they will acquire
a time $t_N$ later, divided by $t_N$.  If we liken the large-scale
fluctuations to harmonic oscillators, they start with zero amplitude and
maximum velocity, and consequently have large time derivatives.

\section{Drift velocities in action space}\label{sec:drift}

We now use the \Nbody\ data to compute the velocities with which stars drift
through action space. This is a more sophisticated and direct probe of
relaxation than $\ex{(\Delta E)^2}$, which is just a global average of the
extent to which stars shuffle parallel to a particular direction in action
space and takes no account of the fact that most shuffles cancel one another:
the distribution function $f(\vJ)$ is unchanged when two stars at similar
action-space locations exchange energy but $\ex{(\Delta E)^2}$ is
incremented.
 
H18 show that in the case of a spherical system, action space, which
is generically three-dimensional with Cartesian coordinates
$(J_r,J_z,J_\phi)$, can be reduced to the two-dimensional space spanned by
$\widetilde\vJ\equiv(J_r,L)$, where $L\equiv J_z+|J_\phi|$ is the magnitude
of the angular momentum vector. The mass density of stars in this space is
$(2\pi)^{3}$ times
\[\label{eq:f2}
f_2\equiv 2Lf,
\]
where $f$ is the full DF, normalised such that $M=(2\pi)^3\int\d^3\vJ\,f$.
H18 present plots of the stellar fluxes in
two-dimensional action space computed both from the BL equation and from
local scattering theory. Their fluxes were normalised to the case of $N=10^5$
stars in a cluster. In our comparisons below we renormalise their fluxes to
clusters of $1000$ stars.

\begin{figure}
\centerline{
\includegraphics[width=\hsize]{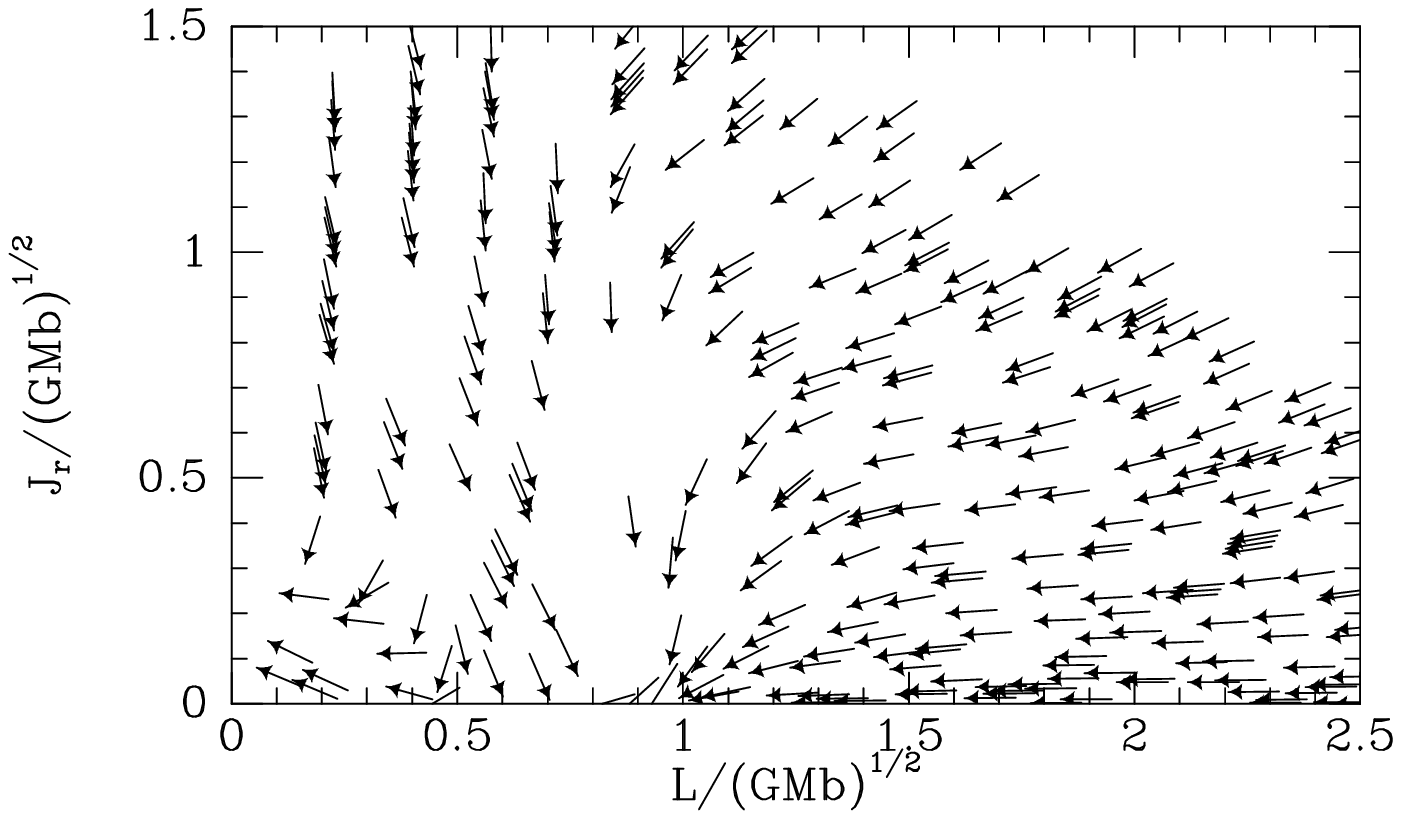}}
\centerline{
\includegraphics[width=\hsize]{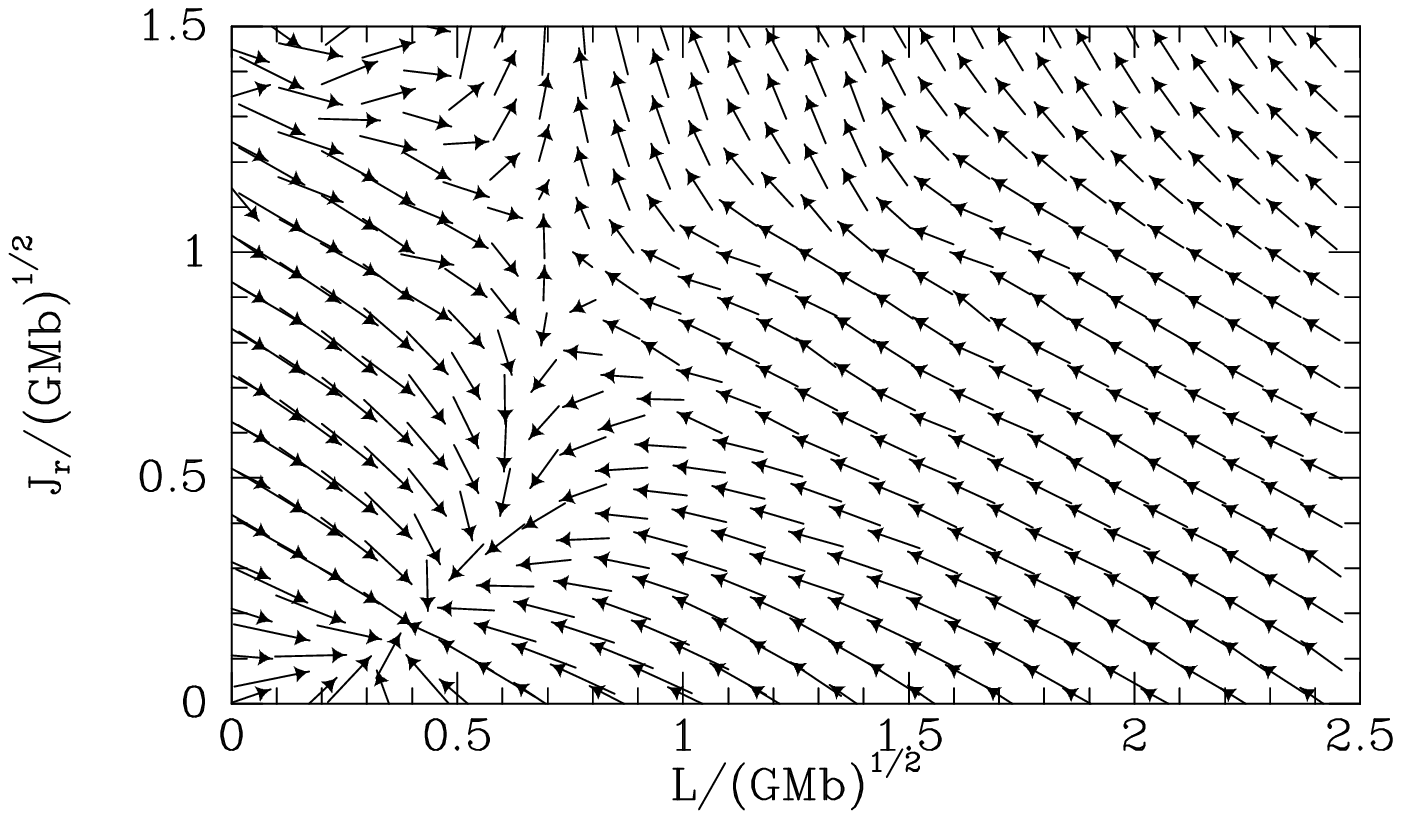}
}
\centerline{
\includegraphics[width=\hsize]{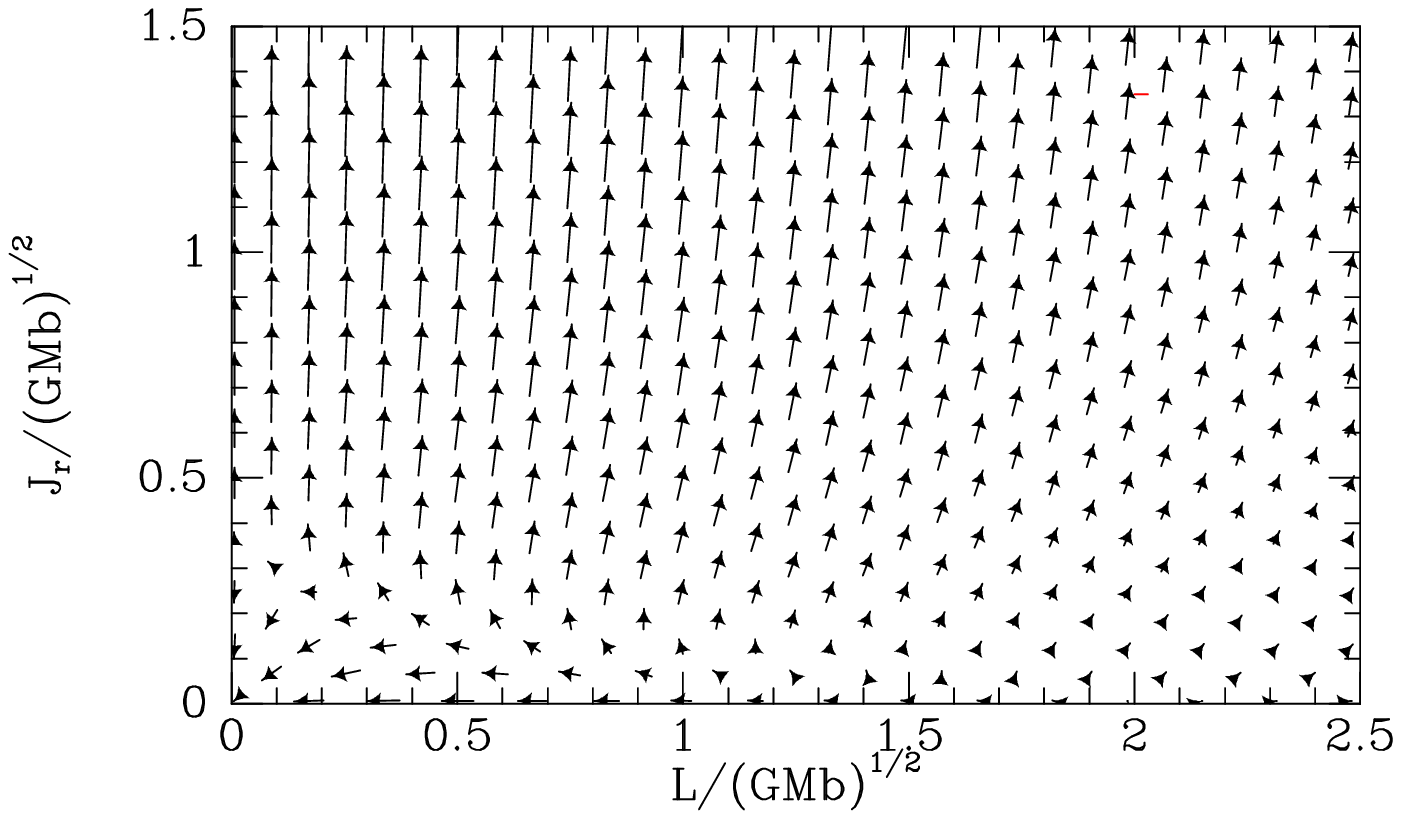}
}
\caption{The drift velocity  through two-dimensional action space obtained
from the BL equation (top), \Nbody\ data via equation (\ref{eq:flux})
(centre) and local scattering theory (bottom). The length of an arrow
represents the magnitude of the velocity using the non-linear scale of
equation (\ref{eq:Vscale}).}\label{fig:flux}
\end{figure}

The flux $\vF$ in the $\widetilde\vJ$ plane is associated with a drift velocity
\[
\vV(\vJ)\equiv{\vF(\vJ)\over f_2(\vJ)}.
\]
From each snapshot it is straightforward to compute each
particle's value of $\widetilde\vJ$ from its energy and angular momentum
because for the isochrone we have an expression for the Hamiltonian
\begin{equation}
H(\widetilde\vJ) = -\frac{(GM)^2}{2[J_r + \fracj12(L + \sqrt{L^2 +4GMb})]^2},
\end{equation}
so we can solve for $J_r(E,L)$.
We divide two-dimensional action space into cells of varying size and we
compute the mean $\bigl\langle{\Delta\widetilde\vJ}\bigr\rangle$ of the
changes in $\widetilde\vJ$ experienced by stars in each cell during the last
half of the simulations, i.e., a time $\Delta t=5t_N$. The drift velocity at
the centre of the cell is taken to be
\[\label{eq:flux}
\vV={\ex{\Delta\vJ}\over\Delta t}.
\]
The centre panel of Fig.~\ref{fig:flux} shows the resulting velocities,
while the top and bottom panels show the velocities that correspond to the
flux vectors computed by H18 from the BL equation and local scattering
theory, respectively. The length $\ell$ of an arrow encodes the speed to the flow
via the non-linear scaling
\[\label{eq:Vscale}
{\ell\over(GMb)^{1/2}}=1.25\times10^{-3}\,{\rm asinh}(V/V_0).
\]
where $V_0=10^{-5} GM/b$. The length of an arrow is then a linear function
of $V$ for $V\ll V_0$ and a logarithmic function for $V\gg V_0$.

The three panels differ considerably from one another. The BL vectors in the
top panel have a very strong tendency to point towards the origin, while
outside the bottom left corner, the local-scattering vectors in the bottom
panel point resolutely upward. The \Nbody\ vectors are something of a
compromise between these two extremes, consistent with the BL and
local-scattering vectors capturing opposite halves of what actually happens
in a cluster. A feature of the \Nbody\ vectors that is captured by the BL
vectors is the tendency to converge on a point at low $J_r$ and non-zero $L$.
However, the point of convergence lies at substantially larger $L$ in the BL
panel than in the \Nbody\ panel. A striking feature of the \Nbody\ panel that
is captured by neither the other panels is a strong tendency for vectors to
align with lines of constant $H$, which run from upper left to lower right.
The \Nbody\ and the BL vectors have similar lengths at all points, while the
vectors from the local approximation are shorter, especially at low $J_r$.
Thus the local approximation materially under-estimates the speed of
relaxation.

Given that H18 included only a handful of resonant pairings, it is inevitable
that the BL vectors differ significantly from the \Nbody\ vectors. The
conflict between the local-scattering and \Nbody\ vectors clearly confirms
that important physics is missed by local-scattering theory.  That physics is
the excitation of system-scale oscillations that is made possible by
self-gravitation. Thus Fig.~\ref{fig:flux} further confirms the essential
message of H18 that the traditional theory of cluster evolution is highly
unsatisfactory.

\subsection{Entropy generation}

In the central panel Fig.~\ref{fig:flux}, the \Nbody\ velocity vectors appear
to converge around the point $\widetilde\vJ=(0.4,0.2)\sqrt{GMb}$. This
convergence implies that in this region $f_2$ is increasing and consequently
the entropy density is decreasing. In this subsection we show that
notwithstanding the local rise of $f_2$, the cluster's entropy is increasing
as it must be.

With $f$ normalised to unit integral through action space, the  entropy of a
cluster in equilibrium is
\[\label{eq:S}
S = - \int \d^3 \vJ \, f \ln(f h^3),
\]
where $h$ is a constant with the dimension of Planck's constant  to ensure that
the argument of the logarithm is dimensionless. We need to apply this formula
to the case of an spherical system, when $f(J_r,L)$. As in H18 we define as a
third coordinate for action space the cosine $\beta$ of an orbit's  inclination
angle, and then we have
that $\d^3\vJ=\d J_rL\d L\d\beta$. Substituting this expression into equation
(\ref{eq:S}) and integrating over allowed values of $\beta$ we obtain
\[
S = - 2\int \d^2\widetilde\vJ \, Lf \ln(fh^3).	
\]
Differentiating with respect to time, we obtain
\begin{align}
\frac{\d S}{\d t} &= - 2\int\d^2\widetilde\vJ\, L{\p f\over\p
t}\left[\ln(fh^3)+1\right]\cr
&= - \int\d^2\widetilde\vJ\, {\p f_2\over\p
t}\left[\ln(fh^3)+1\right],
\end{align}
where the second equality uses equation (\ref{eq:f2}). Differentiation of the
normalisation condition on $f_2$ leads to the conclusion that we can drop the $+1$
from the square bracket. Then we use the BL equation to eliminate $\p
f_2/\p t$ in favour of the flux $\vF$ in $\widetilde\vJ$ space to obtain
\[
{\d S\over\d t}=\int
\d^2\widetilde\vJ\,\nabla_\vJ\cdot\vF\,\ln(fh^3).
\]
The divergence theorem and the fact that in the case of an isotropic system
the DF has the form $f(H)$ allow us to obtain from this
\begin{align}
{\d S\over\d t}&=-\int
\d^2\widetilde\vJ\,\vF\cdot\nabla_\vJ\ln(fh^3)\cr
&=-\int
\d^2\widetilde\vJ\,\vF\cdot\vOmega {\d\ln f\over\d H}.
\end{align}
Given that $\d\ln f/\d H<0$, this equation states that entropy increases if
on average the action-space flux $\vF$ crosses contours of constant $H$
outwards.  

\begin{figure}
\centering
\includegraphics[width = .9\hsize]{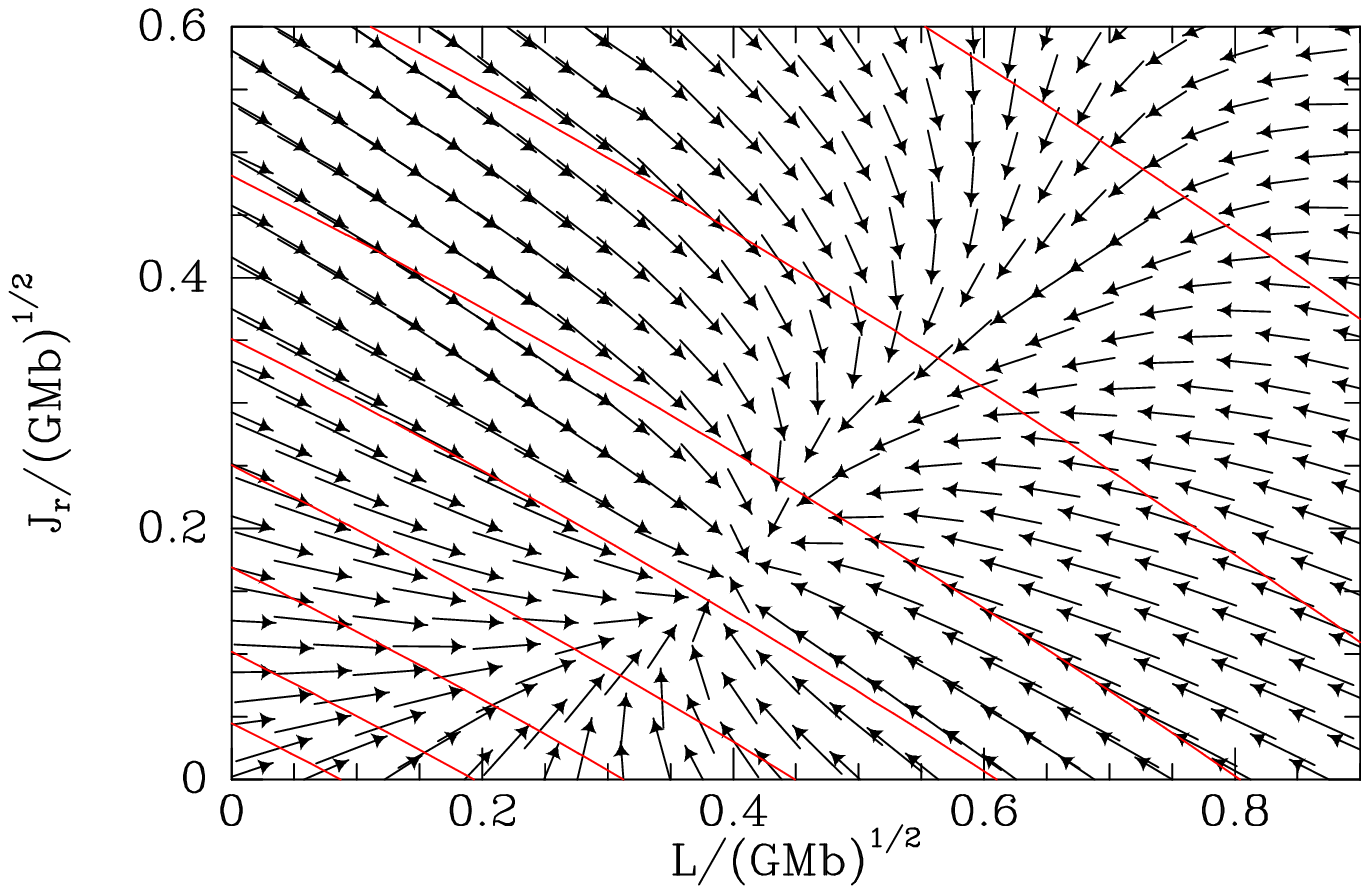}
\includegraphics[width = \hsize]{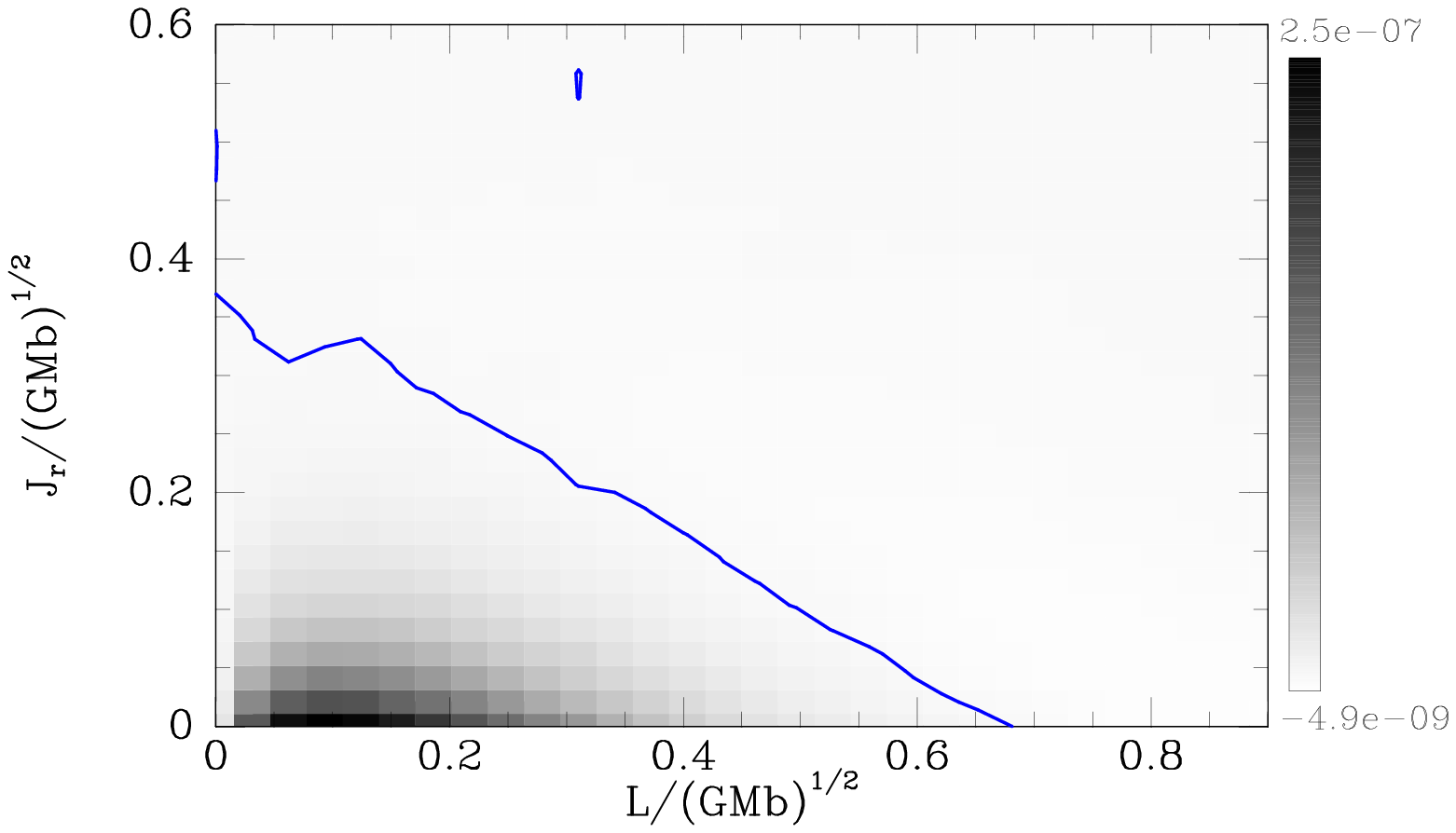}
\caption{Upper panel: the lower-energy part of the central panel of
Fig>~\ref{fig:flux} enlarged and with contours of constant E superposed in
red. Lower panel: the rate of entropy production in this part of action
space. The blue contour divides entropy creation at lower left from entropy
destruction at upper right.}\label{fig:EC}
\end{figure}

The upper panel of Fig.~\ref{fig:EC} combines velocity vectors in action
space with contours of constant $H$ (in red). We see that in much of the
space stars flow approximately along lines of constant $H$. In the lower left
of the figure there is strong upward movement of stars and thus entropy
creation, while in the upper right region stars are moving to lower energy,
so entropy is being destroyed. The lower panel of Fig.~\ref{fig:EC} shows the
rate of entropy generation in a grey-scale plot, with the boundary between
entropy creation and destruction marked by a blue contour. Entropy is
destroyed in most of the depicted part of action space, but the typical
creation rates far exceed the destruction rates, so overall net entropy is
created, as is physically mandatory.

\section{Conclusions}\label{sec:conclude}

A suite of 10\,000 simulations of different realisations of 1000-particle
isochrone clusters demonstrates that the neglect of self-gravity by the
traditional theory of cluster evolution leads to a seriously incomplete
picture of how clusters relax. Over the short durations ($\sim3$ crossing
times) of our simulations, self-gravity amplified the original Poisson noise
by an order-of magnitude. Figs.~\ref{fig:dipole_mag} and \ref{fig:dipole1}
show that the amplification was still some distance from attaining
statistical equilibrium by the end of our simulations, so our estimates of
the importance of self-gravity must be on the low side. An important
implication of this finding is that when a simulations is started by randomly
sampling a DF, it starts from an anomalously regular configuration that would
not occur in nature. Several crossing times are required for the noise level
in the system to settle to a stationary level.

Our experiments were stimulated by the finding of H18 from the BL equation
that self-gravity substantially increases the relaxation rate of clusters.
Their conclusion was in large measure based on the demonstration of strong
excitation by Poisson noise of a cluster's fundamental dipole mode, a result
presaged by the far-sighted work of \cite{Weinberg1993,Weinberg1998}.
Our direct-summation N-body simulations confirm that the dipole mode, in
which the centre moves in opposition to the halo, is powerfully excited. 

We have investigated the extent to which system-scale fluctuations in the
gravitational potential can account for diffusion in energy, which is
traditionally ascribed to the cumulative effect of stars scattering off other
stars. This analysis is tricky because the fluctuations are far from a
stationary random process during our brief simulations. However, 
Fig.~\ref{fig:E} provides strong evidence that these fluctuations are at
least as effective as star-star scattering in causing the energies of stars
to evolve.

Clusters evolve through stars changing their locations in action space. H18
showed that the action space of a spherical system can be reduced to the
two-dimensional space spanned by radial action $J_r$ and the magnitude $L$ of
the angular-momentum vector. Within cells in this space we have determined the
mean drift velocity of stars in the N-body simulations and compared these
with predictions from
the BL equation and local-scattering theory. Fig.~\ref{fig:flux} shows that
the complex action-space flow recovered from the N-body simulations differs
from those predicted from the BL equation on the one hand and
local-scattering theory on the other in ways that suggest that each of the
competing predictions captures roughly a half of the whole picture. The
theoretical predictions differ from one another in that the prediction obtained from the BL
equation by H18 is known to be incomplete, and is in principle capable of
improvement, while that derived from local-scattering theory is definitive and
incapable of modification.

Our final conclusion is, therefore, that local-scattering theory is of very
limited value. With a judiciously chosen value of its free parameter,
$\ln\Lambda$, it can be forced to yield a reasonable value for the relaxation
rate. It also correctly predicts that stars in the cluster's halo gain energy
at the expense of stars in the core. But it is seriously misleading as
regards the basic physics of relaxation and predicts the flow of stars
through action space wrongly. 

It seems then that to continue work on cluster
evolution based on local-scattering theory would be a mistake. Future work
should either rely on N-body simulations, or employ the BL equation.  For the
latter to become a realistic option, one would have to develop a much more
efficient code for the computation of diffusion coefficients than that used
by H18. the extent to which this is possible is unclear, but it is surely
worth a try. 

We employed the direct-summation code \Nbody\ because we were
anxious to exclude any suggestion that our simulations did not handle
star-star scattering accurately. If, as now seems likely, relaxation is
largely driven by system-scale fluctuations, costly direct-summation could be
replaced by a less costly `collisionless' method of force determination, such as tree
summation or Fourier transforms. This is another promising direction for
future work.

\bibliographystyle{mn2e} \bibliography{/u/tex/papers/mcmillan/torus/new_refs}

\appendix\section{Units}\label{app:units}

The isochrone has natural length- and time-scales, $b$ and $(GM/b^3)^{-1/2}$.
Initial conditions for an N-body simulation of an equilibrium system also
define natural length- and time-scales, $r_N$ and $t_N$. Here we determine
the relations between these two systems when the initial conditions are those
of an isochrone.

From the initial conditions one can compute the cluster's potential energy
\[
W\equiv -\fracj12\sum_{i\ne
j}{Gm^2\over|\vx_i-\vx_j|},
\]
 where $m$ is the mass of each of the $N$ stars in the cluster.
For any given value of  $r_N$, we have
\[\label{eq:Wtilde}
W=-
{GM^2\over2r_N}\ex{{1\over|\widetilde\vx-\widetilde\vx'|}},
\]
where $\widetilde\vx\equiv\vx/r_N$ is a dimensionless vector. \Nbody\
rescales the coordinates by choosing $r_N$ such
that the expectation value in equation (\ref{eq:Wtilde}) is unity. Then the
cluster's potential energy is 
\[\label{eq:sN}
W=-{GM^2\over 2r_N}. 
\]

From the initial conditions we can compute the mean-square speed
\[
\sigma^2\equiv{1\over N}\sum_{i=1}^Nv_i^2=\ex{v^2}.
\]
By the virial theorem, $\sigma^2$
is related to $W$ by
\[
2K=M\sigma^2=-W={GM^2\over2r_N}.
\]
With $t_N$ any time, and $\widetilde t\equiv t/t_N$ and
\[
\widetilde\vv\equiv{\d\widetilde\vx\over\d\widetilde t}={ t_N\over r_N}\vv,
\]
dimensionless times and velocities, the mean-square of the dimensionless
velocities is
\[\label{eq:sigmatilde}
\widetilde\sigma^2={GM\over 2r_N^3} t_N^2.
\]
\Nbody\ scales times with $t_N$ chosen such that
$\widetilde\sigma^2=\frac12$, so
\[\label{eq:tN}
t_N=\left({r_N^3\over GM}\right)^{1/2}.
\]

The potential energy of the isochrone is
\[\label{eq:Wisochr}
W=-\left({\pi\over2}-{4\over3}\right){GM^2\over2b}.
\]
Comparing this with equation (\ref{eq:sN}) we find
\[
b=0.2375 r_N.
\]
From equation (\ref{eq:tN}) we have
\[
t_N=8.642\left({b^3\over GM}\right)^{1/2}.
\]
The period of orbits confined to the core is $P=2\pi\surd2 t_{\rm I}\simeq t_N$.

\begin{figure}
\centering
\includegraphics[width=.9\hsize]{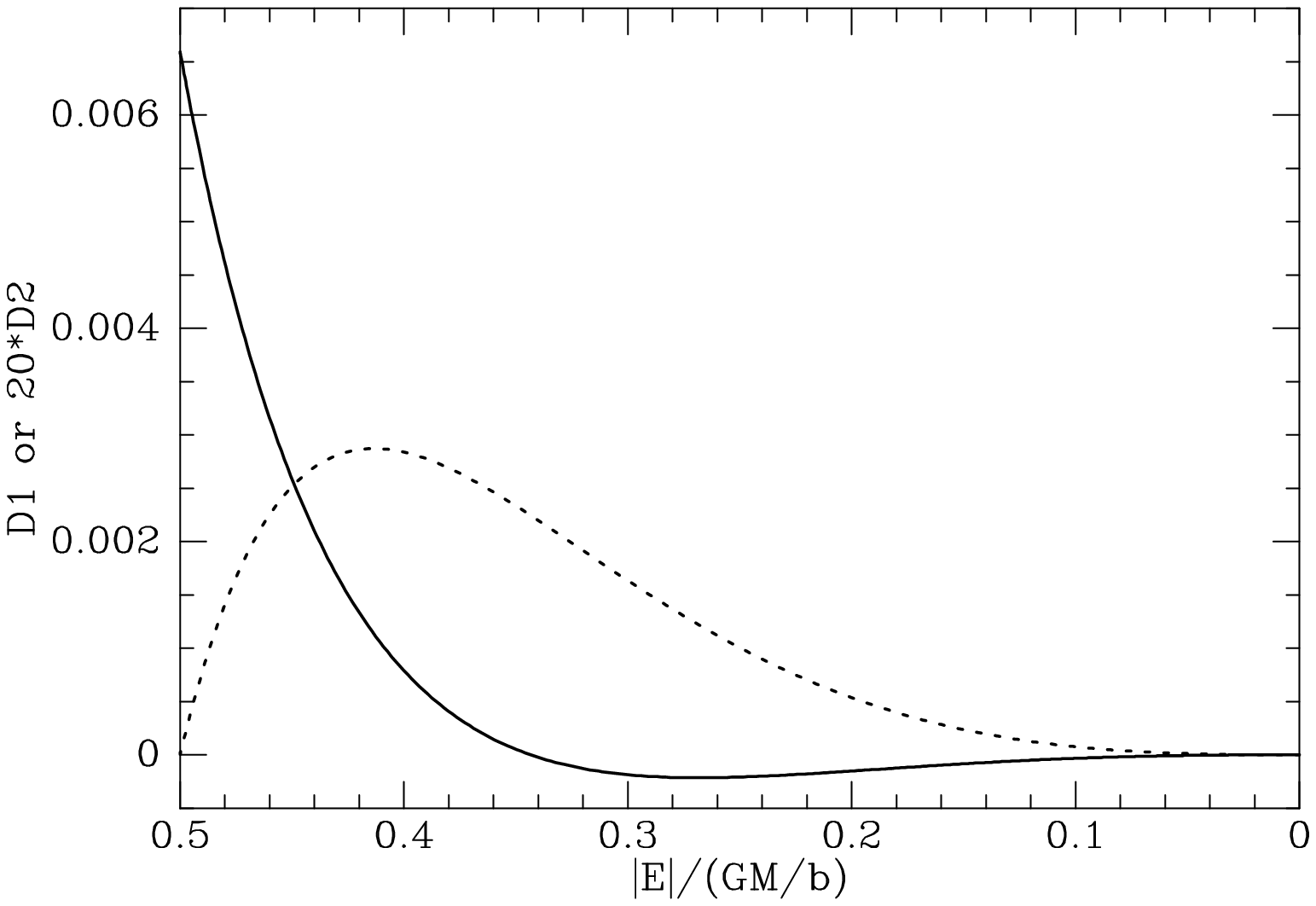}
\caption{The coefficients for diffusion in energy according to
local-scattering theory. The full curve show $\ex{\d E/\d t}/(\sigma^3/b)$ while the
dashed curve shows twenty times $\ex{\d E^2/\d t}/(\sigma^5/b)$, where
$\sigma^2=GM/b$.}\label{fig:dE}
\end{figure}

\section{Diffusion coefficients in energy}\label{app:DE}

Here we list the formulae from \cite{Theuns1996} and \cite{Spitzer1987} from which we
have computed the coefficients that govern diffusion in energy in the
approximation of local scattering.

In a system of total mass $M$ comprising $N$ stars the density of states is
\[
g(E)=(4\pi)^2\int_0^{r_{\rm max}(E)}\d r\,r^2v(E,r)
\]
where $v(E,r)=\sqrt{2(E-\Phi)}$ and $r_{\rm max}(E)$ is the radius at which
$v$ vanishes. A related quantity, the density of states weighted by $v^2$, is
given by
\[
h(E)=(4\pi)^2\int_0^{r_{\rm max}(E)}\d r\,r^2v^3.
\]
With the DF normalised so that $\int\d E\,g(E)f(E)=1$, the diffusion coefficients are 
\begin{align}
\ex{{\d E\over\d t}}&=-{(4\pi GM)^2\over N}\ln\Lambda\biggl[
{1\over g(E)}\int_{-\infty}^E\d E'\,g(E')f(E')\cr
&\qquad-\int_E^\infty\d E'\,f(E')
\biggr]\cr
\ex{{\d E^2\over\d t}}&={2(4\pi GM)^2\over N}\ln\Lambda\biggl[
{1\over g(E)}\int_{-\infty}^E\d E'\,h(E')f(E')\cr
&\qquad+{h(E)\over g(E)}\int_E^\infty\d E'\,f(E')
\biggr].
\end{align}
Fig.~\ref{fig:dE} shows the diffusion coefficients. The first-order
coefficient changes sign from positive to negative as the binding energy
decreases. A useful check on the numerics is provided by the requirement of
energy conservation that $\int\d E\,g(E)\ex{\d E/\d t}=0$.

 \end{document}